\documentstyle[twocolumn,aps]{revtex}

\begin{document}
\draft
\title{Geometry of fully coordinated, two-dimensional percolation}
\author{E. Cuansing, J. H. Kim\cite{kim} and H. Nakanishi}
\address{Department of Physics, Purdue University, W. Lafayette, 
IN 47907 U.S.A.}
\date{\today}

\maketitle

\begin{abstract}

We study the geometry of the critical clusters in fully coordinated percolation
on the square lattice.  By Monte Carlo simulations (static exponents) and
normal mode analysis (dynamic exponents), we find that this problem is in the
same universality class with ordinary percolation {\it statically} but 
not so {\it dynamically}.  We show that there are large differences in the
number and distribution of the {\it interior} sites between the two problems
which may account for the different dynamic nature.

\end{abstract}


\narrowtext
\section{Introduction}
\label{sec:intro}

The geometrical phase transition known as percolation (see, for a review,
Stauffer and Aharony \cite{stauffer94}) is appreciated by many to be an 
elegant and simply defined yet fully featured example of a second order 
phase transition.  A number of variations of the original percolation problem 
were proposed as better models of some physical phenomena in the past.  This 
includes the {\it backbone} percolation for studying electrical conduction 
through random media, {\it polychromatic} percolation for multi-component 
composites, and {\it four-coordinated} bond percolation for hydrogen-bonded 
water molecules.  In particular, Blumberg et al \cite{blumberg80} and Gonzalez 
and Reynolds \cite{gonzales80} studied a random bond, site-correlated 
percolation problem they call four-coordinated percolation on the square 
lattice.  They conclude that this problem belongs to the same universality 
class as the ordinary random percolation with the same set of (static) 
exponents.

In this paper, we revisit a problem in this realm, though not exactly the
same one.  We define {\it fully coordinated percolation} as the site 
percolation problem where only the occupied sites all of whose neighboring
sites are also occupied can transmit connectivity.  Since the random element
is the site, this problem is slightly different from the bond problem
referred to above.  Thus, after generating a random site configuration with
the independent site occupation probability {\it p}, we only select those
occupied sites with all 4 neighbors also occupied on the square lattice and
study the clusters formed by nearest neighbor connections among those sites.
It should be noted that this problem is distinct from the so-called
bootstrap percolation (see, e.g., \cite{privman91}) where sites of less
connectivity are iteratively removed. In our problem, no iterative procedures
are involved; rather, sites of less than full connectivity are marked first
and then all of them removed at one time.

This problem arose in the context of studying the vibrational properties of
fractal structures tethered at their boundaries \cite{mukherjee96,mukherjee98}.
In that problem, scaling was observed in the normal mode spectrum whose
origin may lie in the ratio of 2 length scales, one of which 
is the size of highly connected regions of a cluster.  In this context, we 
have embarked on revisiting the characteristics of randomly generated, but 
highly connected geometrical structures.

In the next section, we summarize the Monte Carlo and finite size scaling
analyses of the static critical properties of fully coordinated percolation.
In Section 3, we discuss the normal modes of the {\it transition probability
matrix} for tracer diffusion on the structure using the methods of Arnoldi
and Saad (see, e.g., \cite{nakanishi94}).  Then in Section 4, we describe the 
classification of the cluster sites into external boundary, internal boundary,
and interior ones and using these to show the major distinctions between
the critical clusters of ordinary and fully coordinated percolation.  We
summarize the results in the final section.

\narrowtext
\section{STATIC CRITICAL BEHAVIOR}
\label{sec:static}

To determine the static critical behavior of fully coordinated percolation we
first performed Monte Carlo simulations on a square lattice in two dimensions.
Each site is occupied with probability $p$
independently and subsequent fully coordinated sites are marked and their
connectivity searched.  Lattice sizes of $L^{2}$ where $L = 256$, $512$, 
$1024$, and $2048$ were constructed.  For each lattice size we further made 
a thousand realizations wherein a different random number seed was used on 
every run.  The unnormalized susceptibilities, i.e.,
$\Xi(L) = \sum_{s}^{'} s^{2} \hat{n}_{s}$ where $\hat{n}_{s}$ is
the number of clusters of size $s$, are calculated on each run and
are then summed at the end of the thousand realizations.  The average 
susceptibilities $\chi$ are calculated by dividing the sum
by the number of realizations and the lattice size.
The prime on the summation indicates the fact that the contribution of
the largest cluster to $\chi$ near and above what we perceived to be the
critical probability $p_{c}$ has been subtracted as usual \cite{stauffer94}.  

In Fig. 1 we plot the average susceptibilities against the probability
$p$ for the corresponding lattice sizes.  The data correspond to the values
of $L = 256$, $512$, $1024$, and $2048$ from the lowest to highest.
We can see that the effects due to the finite sizes of the lattices are 
exhibited clearly.  In particular, there are well-defined peaks which 
scale with lattice sizes as
\begin{equation}
\chi(p_{max}, L)~\sim~L^{\gamma/\nu}
\label{equ:peaks_scaling}
\end{equation}
where the known exact value of $\gamma/\nu$ for the ordinary percolation is
$\frac{43}{24} \sim 1.7917$.  To demonstrate the precision of our 
calculations, we plot $\chi(p_{max}, L)$ against the corresponding lattice 
sizes in the inset of Fig. 1.  Notice that the data follow an excellent 
power law, leading to a least squares fit of $\chi(p_{max}, L) \sim 
L^{1.7911}$.  The value of $\gamma$ found is identical with the ordinary 
percolation value to within about $0.03\%$.  This result confirms previous 
work\cite{blumberg80,gonzales80} stating that fully coordinated percolation and
ordinary percolation belong to the same {\em static} universality class. 

The critical behavior of susceptibility is known to scale as \cite{stauffer94}
\begin{equation}
\chi(p, \infty)~\sim~|p~-~p_{c}|^{-\gamma}
\label{equ:chi_scaling}
\end{equation}
where for ordinary percolation $\gamma_o = \frac{43}{18} \sim 2.3889$.  Notice 
however that in Fig. 1 the peaks are very near $p = 1.0$.  This would provide
data to the right of the peaks in only a small probability interval.  In our
simulations, we would therefore use $\chi$ only to the left of the peaks.

Since the scaling relation in Eq.\ (\ref{equ:chi_scaling}) is expected only
for infinite lattices, we use only the data taken from $L = 2048$ to test it.
Since there are two unknowns in Eq.\ (\ref{equ:chi_scaling}), we first choose a
particular $p_c$ and make a fit to see what value of $\gamma_{exp}$ is 
obtained.  If we choose $p_c = 0.886$ we get $\gamma_{exp} = 2.4004$.  The 
correlation coefficient, $|R|$, for this fit is $0.99999$.  The discrepancy 
between $\gamma_{exp}$ and $\gamma_o$ is around $0.481\%$.  Choosing
$p_c = 0.8858$ we obtain $\gamma_{exp} = 2.3864$.  The discrepancy this time
is around $0.10\%$ and $|R| = 1$.  So we have an exact fit for this value of
$p_c$ and the $\gamma_{exp}$ found is very close to the $\gamma_o$.  Choosing
$p_c = 0.885$ we obtain $\gamma_{exp} = 2.3302$ with an $|R| = 0.99999$.  The
discrepancy for this value of $p_c$ is $2.46\%$.  Fits done with $p_c$ between
$0.885$ and $0.886$ gave $|R| = 1$; however, the $\gamma_{exp}$ found when
$p_c = 0.8858$ gave the closest value to $\gamma_o$.  This allows us to 
conclude that $\gamma$ for fully coordinated percolation is the same as that
for ordinary percolation while also giving an estimate for the value of
$p_c$ close to $0.8858$.  (We will state the experimental
uncertainty for $p_{c}$ after all our analyses are presented.)

From the fit done to examine the scaling in Eq.\ (\ref{equ:peaks_scaling}) we
could further conclude that $\nu$ for fully coordinated percolation should
be the same with that for ordinary percolation.  This again confirms the
statement that fully coordinated percolation is in the same {\em static}
universality class as ordinary percolation. Another universal constant often
used to characterize ordinary percolation is the amplitude ratio
$C_{+}/C_{-}$ of susceptibility $\chi$ (whose value is about 200 in $d=2$
\cite{stauffer94}). In fully coordinated percolation,
this quantity is unfortunately difficult to calculate accurately because the
critical region for $p > p_c$ is very small (see below). When we constrain
the exponent $\gamma$ to be close to $\gamma_o$ and use the $p_c$ estimated
in this work, however, we find that $C_{+}/C_{-}$ is of $\vartheta(10^2)$,
which is consistent with the above observation as well.

The contribution of the largest cluster to the susceptibility is not
significant when $p < p_{c}$.  However when $p \sim p_{c}$~a significant
number of sites will belong to the largest cluster and when $p > p_{c}$
the largest cluster is dominant in the whole lattice.
The average susceptibility contribution due to this largest cluster
is $\chi_{1} = \sum s^{2}_{max}/(L^{2} N)$, where the summation is over
$N = 1000$ realizations and $s_{max}$ is the size of the largest cluster.
The fractal dimension, $d_f$, can be obtained from $s_{max}$ by
\begin{equation}
s_{max}(p_c)~\sim~L^{d_f}
\label{equ:frac_dim}
\end{equation}
where $\overline{s}_{max}(p_c)$ is the mean size of the largest cluster
at $p_c$.  $\chi_{1}$ should therefore scale as
\begin{equation}
\chi_{1}~\sim~L^{2 d_f~-~2}.
\label{equ:large_scaling}
\end{equation}
For ordinary percolation on a two dimensional lattice (see, e.g.,
\cite{stauffer94}), it is known that $d_f = 91/48$ and
$y \equiv 2 d_f - 2 \sim 1.7917$.  For fully coordinated percolation, the
scaling in Eq.\ (\ref{equ:large_scaling}) have two unknowns, $p_c$ and $d_f$.
Similar to what we have done when examining the scaling in 
Eq.\ (\ref{equ:chi_scaling}), we choose trial values for $p_c$ and then
perform a least squares fit to obtain the corresponding $y = 2 d_f - 2$.
By looking for the range of trial $p_c$ that maximizes the regression
coefficient $|R|$, we arrive at an estimate of $p_c$ to be close to $0.8845$
where $|R| = 1$ and $y = 1.7855$. The variation of $|R|$ is about 2 parts in
$10^5$ if $p_c$ is varied by 0.0002, always with less than 1\% deviation
from the ordinary percolation value of $y$.  From these results
we conclude that $d_f$ for fully coordinated percolation is the same as that 
for ordinary percolation as well as the estimate of about $0.8845$ for $p_c$.

In addition to the above, we have also performed the scaling analysis of
the quantity $\chi_1 (p,L)$, as both $p$ and $L$ are varied, in the form of
\begin{equation}
\chi_{i1}~=~L^{2 d_f~-~2}~g(|p~-~p_c|^{\nu}~L)
\label{equ:large_outside_scaling}
\end{equation}
where $g(|p - p_c|^{\nu}~L)$ is a scaling function. Using the exactly known
ordinary percolation values of the exponents $d_f$ and $\nu$ (as they have
been shown to be the same for fully coordinated percolation above),
we obtained the maximum data collapsing in the range of
$0.884 < p_c < 0.885$.

Independent of the above analyses based on the fully coordinated clusters
obtained by Monte Carlo simulations of fixed-sized square grids, we have
also performed Monte Carlo simulations by growing fully coordinated
clusters starting from a seed site using a variant of the {\em breadth-first
search} algorithm \cite{nakanishi94}. This latter approach has an advantage
that there is no obvious finite-size effects and that statistics taken
while a cluster is still growing represents a {\em partial sum} automatically.
That is, we start growing such clusters 10,000 times at each of $p = 0.880$,
0.884, 0.885, and 0.890, and keep track of how many of them are still
growing at predetermined intervals of size ($2^n$ where $n = 1$, 2, ... , 15
in our case). This number, say, $N_s$ represents the partial sum

\begin{equation}
N_s /N_1 = \frac{1}{p} \sum_{s' \geq s} s' n_{s'}
\end{equation}

Since the normalized number of size-$s$ clusters, $n_s$ scales as
$s^{-\tau} f(\epsilon s^{\sigma})$ where $\tau =187/91$ and $\sigma = 36/91$,
we expect that $N_s$ scales as
\begin{equation}
N_s s^{\tau -2} \sim \hat{f}(\epsilon s^{\sigma})
\end{equation}
near $p_c$ and for large $s$. In particular, at $p_c$, this quantity
should be constant independent of (large) $s$. The numerical results are
shown in Fig.~2, where the data correspond, from highest to lowest, to
$p=0.890$, 0.885, 0.884, and 0.880. The horizontal dashed line drawn
to guide the eye makes it clear that data for $p=0.885$ best approximates
a horizontal line as $s \rightarrow \infty$, suggesting that a good estimate
of $p_c$ would be 0.885.

We now consider all the above results together. The results from scaling in
Eq.\ (\ref{equ:chi_scaling}) indicate a range $0.885 - 0.886$, and
those from scaling in Eq.\ (\ref{equ:large_scaling}) indicate $0.8844 - 0.8847$,
while those from scaling in Eq.\ (\ref{equ:large_outside_scaling}) hints
$p_c$ to be in the $0.884 - 0.885$ interval. Another result that could also
be used are the values of $p$ for the peaks in Fig.~1, which
vary from $0.8841$ to $0.8844$ (with the peak for the largest grid
$L = 2048$ occurring at $p_{peak} = 0.8844$).  Combining all these 
results, our final estimate is $p_c = 0.885 \pm 0.001$.

\narrowtext
\section{DYNAMIC CRITICAL BEHAVIOR}
\label{sec:dynamic}

By dynamic critical behavior here we simply mean the asymptotic long-time
behavior of diffusion taking place on an incipient infinite cluster of
fully coordinated percolation, or its equivalent {\em scalar} elastic
behavior. This represents the simplest kind of dynamics associated with
these complicated geometrical objects and is mainly reflected in
the two {\em dynamic} critical exponents called $d_s$
(spectral dimension) and $d_w$ (walk dimension).

It is well known that the return-to-the starting point probability of
the random walk, $P(t)$, in the long-time limit obeys the power law,
\begin{equation}
P(t) \sim t^{-d_s/2} , 
\end{equation}
where $d_s$ is the spectral dimension of the walk. In a fractal medium,
$d_s$ is less than the space dimension $d$, because the progressive
displacement of the random walker further from the starting point is
hampered by its encounter with the irregularities of the medium at all 
scales. Thus, $d_s$ is expected to be greater for environments that
provide higher connectivity at large length scales, independently of
the fractal dimension itself which is mainly the measure of the overall
{\em size} scales or {\em how many} sites are connected,
not {\em how well} those sites are connected to each other.

For media with long-range loops, $d_f$ (fractal dimension), $d_s$,
and $d_w$ are not independent but are expected to obey the well-known
Alexander and Orbach scaling law \cite {alexander} 
\begin{equation}
\label{eq:ao}
d_s = 2d_f/d_w  \;. 
\end{equation}
For this reason, we only calculate $d_s$ here though
both of $d_s$ and $d_w$ can be conveniently calculated by numerically
studying the {\em transition probability matrix} {\bf W} which represent
the random walk on a specific fractal medium. Our calculation in this work
is only one aspect of such an analysis: finite size scaling of the 
dominant non-trivial eigenvalue which describes the longest finite time
scale of the Brownian process. This approach has already been described
in detail elsewhere \cite{mukherjee94}, and thus we merely state the main
feature and then immediately report our specific numerical results.

The matrix {\bf W} is constructed from the elements $W_{ij}$ being equal
to a hopping probability per step (equal to $\frac{1}{4}$ here) for 
available nearest neighbor sites $i$ and $j$. For each neighbor site
which is not present, a probability of $\frac{1}{4}$ is added to the
probability for not taking a step for one time period - this is called
the {\em blind ant} rule. Many large matrices {\bf W} are obtained
by Monte Carlo simulation (by growing a fully coordinated
percolation cluster from a seed site and stopping the growth when a
predetermined desired size is reached) and their largest eigenvalues
are numerically obtained by the so-called {\em Arnoldi-Saad} method.
The dominant non-trivial eigenvalue $\lambda_1$ is the largest eigenvalue
just below the stationary eigenvalue 1 and it is known to satisfy
the following finite size scaling law:

\begin{equation}
\label{eq:fsfords}
| \ln \lambda_1 | \approx 1-\lambda_1 \sim S^{-2/d_s} \;.
\end{equation}

Shown in Fig.~3 are our results from such an analysis.
We have generated at least 1000 independent realizations of the underlying
fully coordinated percolation clusters for sizes $S = 1250$, 2500, 5000,
10000, and 20000 at each of the three nominal probabilities
$p=0.883$, 0.885, and 0.887, and numerically obtained $\lambda_1$
for each cluster.  The main part of Fig.~3 shows the data from $p=0.885$,
the value shown to be closest to $p_c$ in this work. The figure 
shows an excellent power law fit (regression coefficient of -0.99993)
to Eq.(\ref{eq:fsfords}) with the exponent $2/d_s = 1.486 \pm 0.01$.
This power translates to $d_s = 1.346 \pm 0.011$ which is close to but
definitely larger than the corresponding ordinary percolation value
of $d_s^{(O)} = 1.30 \pm 0.02$ estimated by many independent calculations.
(See, e.g., \cite{nakanishi94}.) For comparison, a {\em loopless} variant
of percolation has exactly the same static exponents as ordinary percolation
but has $d_s \approx 1.22$ in two dimensions, about twice as much deviation
from ordinary percolation in the opposite direction as the present
fully coordinated percolation problem. \cite{loopless}

In the inset for Fig.~3, we show a normalized $\lambda_1$ by plotting
$(1-\lambda_1) S^{2/d_s^{(o)}}$ where the circles are for $p=0.887$,
squares for $p=0.885$, diamonds for $p=0.883$ and the solid line is
a horizontal line (for ordinary percolation) to guide the eye. In all
cases, the standard errors of the mean for each set of data are
substantially smaller than the size of the symbols used in the figure.
The distribution of $\lambda_1$ in each case appears to be Gaussian
with the standard deviations scaling in the same way as the means.

>From these results, we conclude that, though the numerical differences
are small, there is a high likelihood that the fully coordinated
percolation clusters are significantly different from the ordinary
percolation counterparts even at long length scales. In the next section,
we show that this analysis is vindicated by exposing one dramatic
difference in the cluster morphology which will not be obvious
to an uncritical observer.

\narrowtext
\section{CLUSTER GEOMETRY}
\label{sec:geometry}

In this section we examine the geometry of fully coordinated percolation
clusters more closely. First, we present Fig.~4 which show in grey scale
the sites of (a) fully coordinated percolation cluster and (b) ordinary
percolation cluster at respective $p_c$. The overall visual impression
is that they are very similarly shaped even down to the details of 
the boundaries and internal holes. Their shapes are also essentially
independent of the underlying lattice anisotropy (as Fig.~4(a) has actually
been rotated by 45 degrees with respect to the coordinate axes of the
square lattice). However, the number and distributions of the especially
dark points are evidently quite distinct in (a) and (b). They
cluster more and are much more abundant in (a) than in (b). These sites
are actually the {\em interior} or fully coordinated sites in the internal
part of the cluster. The remaining sites (shaded grey) are either
the external {\em hull} sites or internal boundary sites.

In Fig.~5 quantitative examination is made on the different classes of
sites of the two kinds of clusters. In the main part of the figure,
the average numbers of two kinds of sites are shown, {\em interior}
(diamonds for fully coordinated percolation, crosses for ordinary percolation),
and {\em external} (squares for fully coordinated percolation, plusses
for ordinary percolation). It is clear that the interior sites are more
than 3 times abundant in fully coordinated percolation as is visually
suggested by Fig.~4. Though this is primarily a local effect due to the
full coordination rule, they do have a multiplicative effect at long-range
connectivity and thus may well be the source of the small difference in the
value of $d_s$.

Of course just the fact that there are more than 3 times as many
interior sites (and correspondingly, much fewer {\em hull} sites)
in fully coordinated percolation must have quantitative
consequences (even if not qualitative) for any process on the cluster
which depend on degrees of connectivity rather than just on the number of
connected sites. An example of the effect of the different numbers of
the hull sites may be in oxidation or catalysis of a material through
the external embedding phase or even a irregularly shaped breakwater
in the form of the external boundary of a percolation cluster.
The {\em external} sites are those which are sometimes called
{\em hull} sites and they are known to scale with an exact exponent
in ordinary percolation as

\begin{equation}
N_{hull} \sim S^{d_h /d_f}
\end{equation}
where $d_h^{(o)} = 7/4$ and thus the exponent is $84/91 = 0.923...$
for ordinary percolation. Since this is less than 1, these sites comprise
less and less fraction of the cluster as $s \rightarrow \infty$ and the
remaining sites (i.e., {\em interior} and {\em internal boundary} sites)
eventually dominate the whole cluster. This is already evident from the
greater slopes close to 1 for the {\em interior} sites in Fig.~5. The
linear regression fits for the {\em hull} sites in Fig.~5 indicate the
slopes of about 0.913 for ordinary percolation and 0.922 for fully coordinated
percolation with essentially perfect fits, again reinforcing the conclusion
that they show the same {\em static} critical behavior.

\narrowtext
\section{SUMMARY AND CONCLUSION}
\label{sec:summary}

In summary, we have studied both static and dynamic critical behaviors
associated with a model of the highly connected regions of a disordered
cluster. The model is a {\em site} variant of the four-coordinated percolation
\cite{blumberg80,gonzales80} on the square lattice we call
{\em fully coordinated percolation}. While the bond version was studied
for {\em static} critical behavior, neither bond nor site version was
previously studied for the dynamic behavior to the best of our knowledge.
We have used various methods such as Monte Carlo simulations, finite-size
scaling and Arnoldi-Saad approximate diagonalization of large random matrices
for this purpose.

Though all indications are that the static behavior of this model is exactly
the same as the ordinary percolation (as previous work suggested), the dynamic
behavior shows a small but significant difference in the values of the
universal critical exponents. We have looked for the cause of this 
difference and found a three-fold increase in the number and significantly
enhanced clustering of the interior sites (i.e., those not on the exterior
or internal boundaries) and the associated decrease in the number
of boundary sites.  Thus, although the deviations from ordinary percolation
in terms of the values of the dynamic critical exponents are not large,
there will be rather significant differences in any processes that depend
sensitively on those numbers. Possible examples of such processes include
the oxidation of a material through the external embedding phase and 
the vibrational normal modes with boundary conditions such as {\em clamping}
or {\em tethering} of the external boundaries (through the contrast in
elastic constants of embedding and embedded materials, for example).

\acknowledgements
One of us (JHK) wishes to thank the Purdue University Department of Physics
for the hospitality during his visit there when part of the work was done.
We are also grateful to D. Stauffer and R. Ziff for insightful remarks.

%
%

\begin{figure}
\caption{Susceptibilities $\chi$ for fully coordinated percolation
on finite square lattices of size $L^2$ are shown against probability $p$,
where crosses, diamonds, squares, and circles are for $L=256$, 512, 1024,
and 2048, respectively.
The inset shows the finite-size scaling of the peaks of these curves by
plotting the maximum $\chi$ vs. $L$ on a log-log scale.}
\end{figure}

\begin{figure}
\caption{The scaled partial sum $N_s$ is plotted against $s^\sigma$ for
$p = 0.890$ (cross), 0.885 (circle), 0.884 (triangle), and 0.880 (plus).
The dashed line is a horizontal line to guide the eye, showing that the
data for $p=0.885$ best fits the horizontal slope for large cluster sizes.}
\end{figure}

\begin{figure}
\caption{The scaling of the largest nontrivial eigenvalue $\lambda_1$ of
the Markov chain analysis is shown by plotting $1-\lambda_1$ against
the cluster size $s$ on a log-log scale. The inset shows a normalized
$\lambda$ (see text) for $p=0.887$, 0.885, and 0.883 (from top). The solid
line is horizontal, corresponding to the slope for ordinary percolation.}
\end{figure}

\begin{figure}
\caption{(a) A typical fully coordinated percolation cluster at its $p_c$,
and (b) the corresponding critical ordinary percolation cluster. The dark
points are the {\em interior} sites of full coordination and the grey shaded
sites are either on the external hull or internal boundaries.}
\end{figure}

\begin{figure}
\caption{The numbers of {\em external} sites (squares and plusses) and
those of {\em interior} sites (diamonds and crosses) are scaled against
the cluster size $s$. The solid (fully coordinated percolation) and
dashed (ordinary percolation) lines are the linear least squares fits.
In the inset, we show the number of external sites divided by
$s^{d_h^{(o)} /d_f^{(o)}}$, showing that the population of the external
{\em hull} sites of fully coordinated and ordinary percolation scale
in the same way.}
\end{figure}

%
%

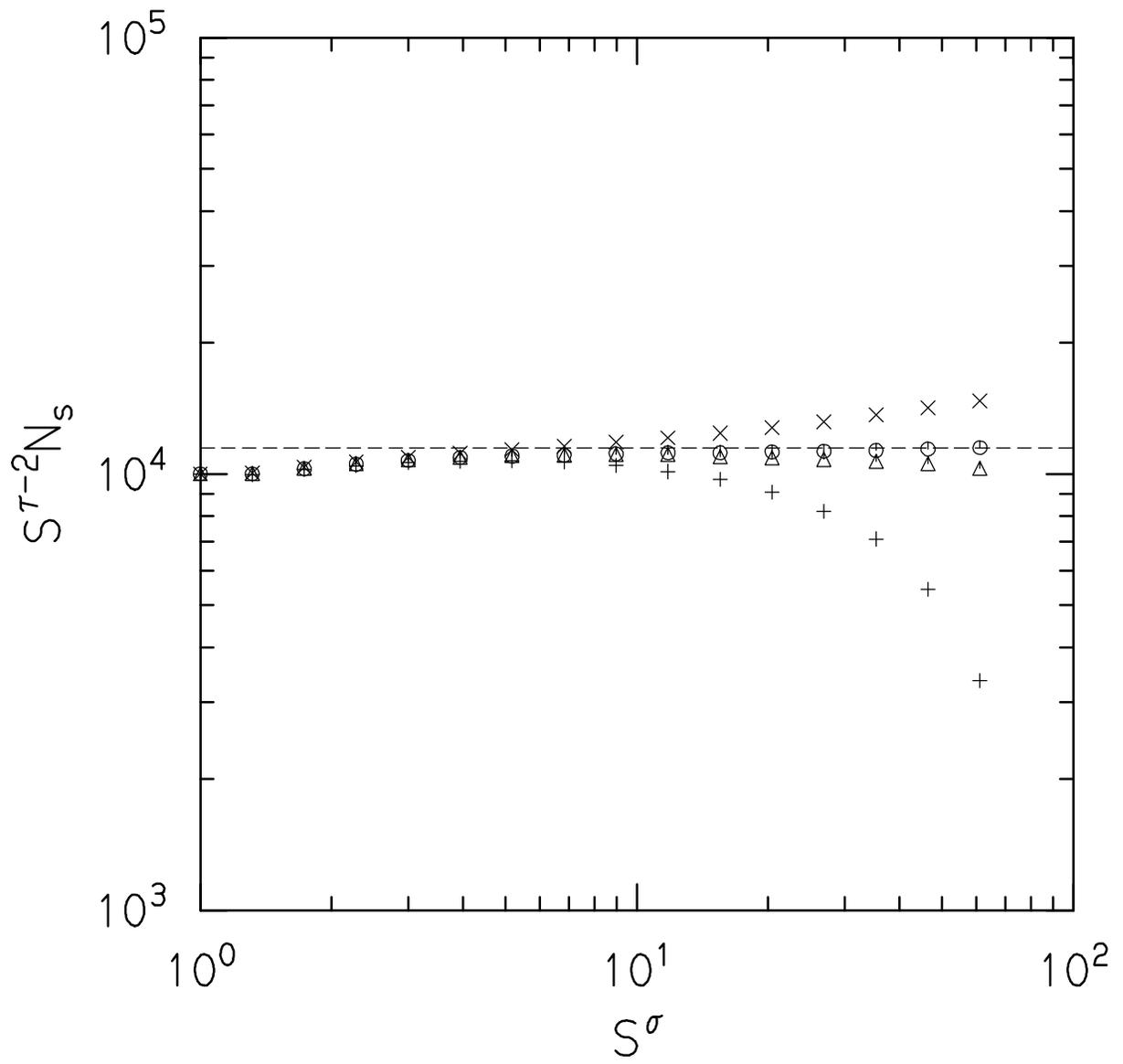
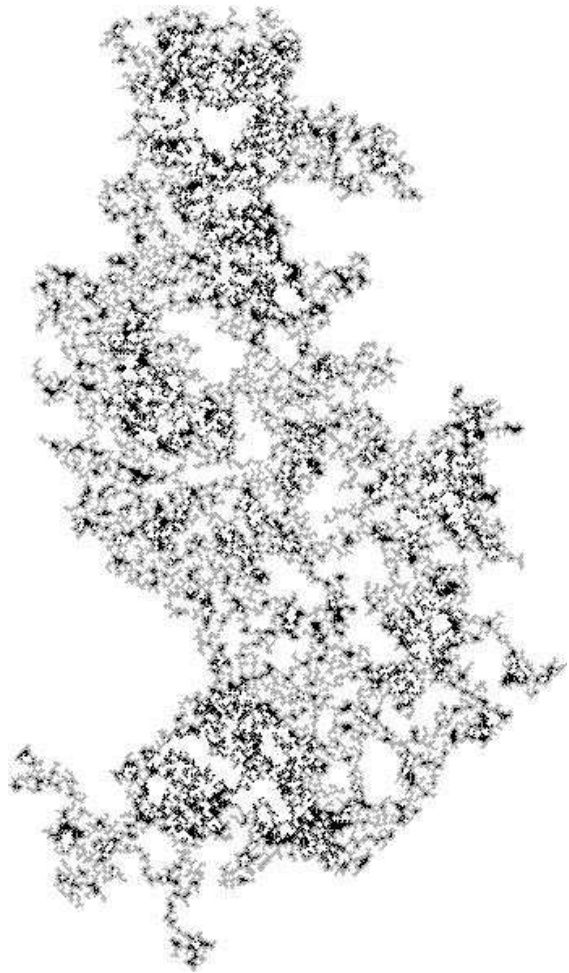
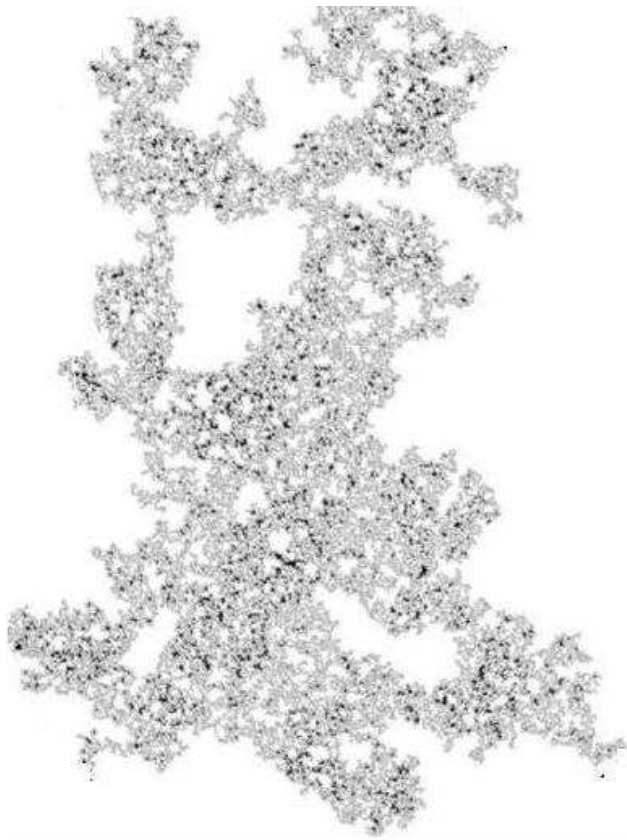
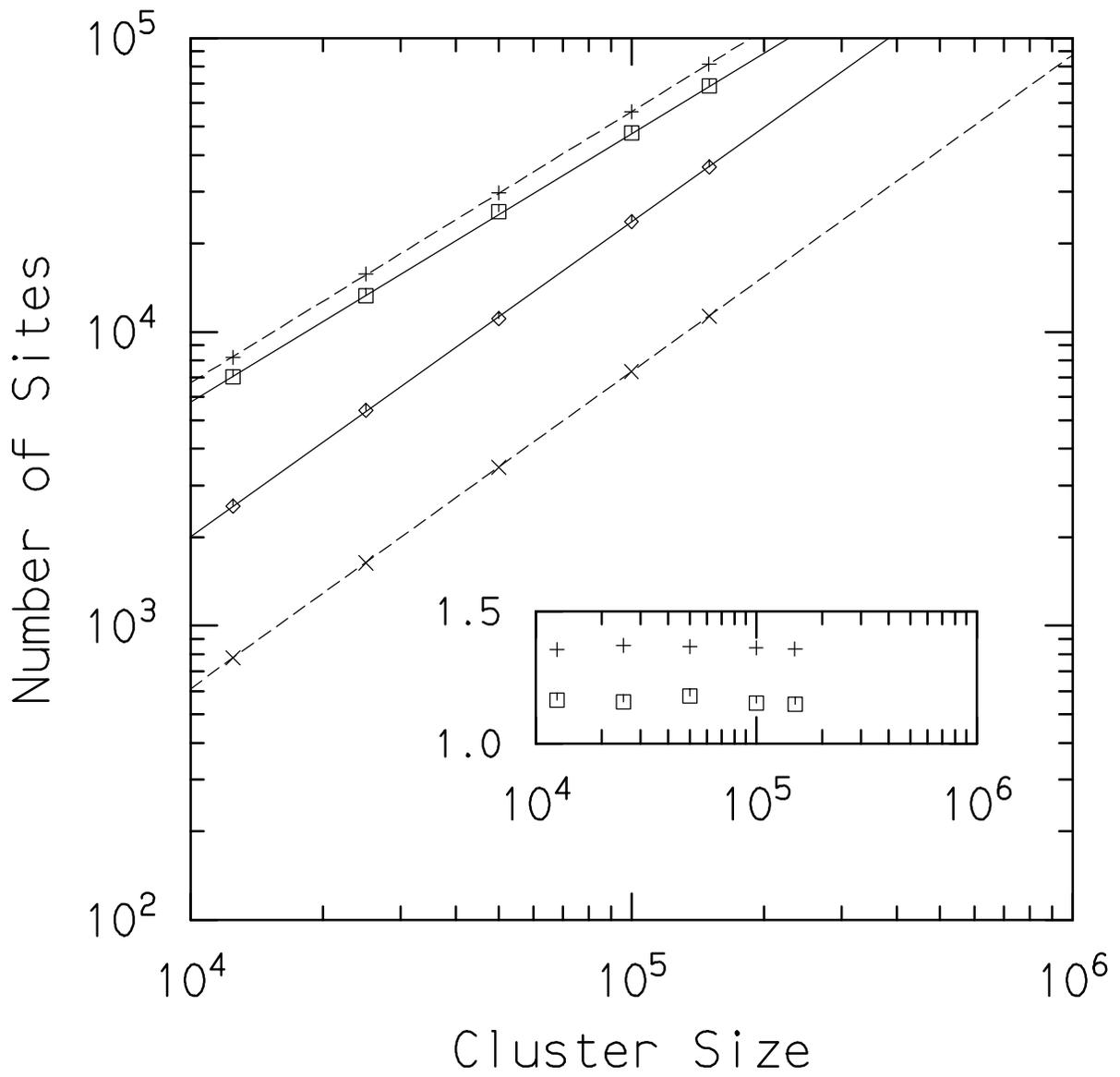
\end{document}